\documentclass[aps,showpacs,showkeys,superscriptaddress,nofootinbib,preprint,11pt]{revtex4}
\usepackage{graphicx}% Include figure files

\begin{document}

\title{A-dependence of weak nuclear structure functions}

\author{H. Haider}
\affiliation{Department of Physics, Aligarh Muslim University, Aligarh-202 002, India}

\author{I. Ruiz Simo}
\affiliation{Dipartimento di Fisica, Universit\'a degli studi di Trento
Via Sommarive 14, Povo (Trento)
I-38123, Italy}

\author{M. Sajjad Athar}
\affiliation{Department of Physics, Aligarh Muslim University, Aligarh-202 002, India}

\begin{abstract}
Effect of nuclear medium on the weak structure functions $F_2^A(x,Q^2)$ and $F_3^A(x,Q^2)$ have been studied using charged current (anti)neutrino
deep inelastic scattering on various nuclear targets. Relativistic nuclear spectral function which incorporate
Fermi motion, binding and nucleon correlations are used for the calculations. We also consider the pion and rho meson cloud contributions calculated from a 
microscopic model for meson-nucleus self-energies. Using these structure functions, $F_i^A/F_i^{proton}$ and $F_i^A/F_i^{deuteron}$(i=2,3, A=$^{12}C$,
 $^{16}O$, $CH$ and $H_{2}O$ ) are obtained.  

\end{abstract}

\maketitle
\section{Introduction}
MINER$\nu$A at Fermi lab is measuring neutrino nucleus cross sections in various nuclear targets in the energy region of 1-20 GeV. Among the various goals,
one of the aim is to study structure functions in the deep inelastic scattering processes.  In this work, we have studied nuclear medium effects 
on the weak structure functions $F_2^A(x,Q^2)$ and $F_3^A(x,Q^2)$, using a relativistic nucleon spectral function that describes the momentum distribution of nucleons in
the nucleus. We define everything within a field-theoretical
approach where nucleon propagators are written in terms
of this spectral function. The spectral function has been
calculated using Lehmann's representation for the relativistic
nucleon propagator and nuclear many-body theory is used to
calculate it for an interacting Fermi sea in nuclear matter. A
local-density approximation is then applied to translate these
results to finite nuclei~\cite{prc84,prc85}.
 We have also taken target mass correction and shadowing effects into account. The details of the model are given in
Refs. \cite{prc84,prc85,marco1996}.\ This paper is aimed to study the nuclear dependence of the weak structure functions $F_2^A$ and $F_3^A$ 
in the charged current neutrino/antineutrino induced deep inelastic reactions using $^{12}C$, $^{16}O$, $CH$ and $H_{2}O$  nuclear targets.
We have obtained proton and deuteron structure functions and the ratio $F_i^A/F_i^{proton}$ and $F_i^A/F_i^{deuteron}$(i=2,3, A=nuclear target) 
to see the nuclear target dependence on the weak structure functions. In our numerical calculations the deuteron structure functions
have been obtained using the same formalism as for the nuclear targets but performing the convolution with the deuteron wave function squared
instead of the nuclear spectral function. 

\section{Formalism}
In the nuclear medium the expression for the differential scattering cross section is written as \cite{prc84}:
\begin{equation} 	\label{cross_nuclear}
\frac{d^2 \sigma_{\nu,\bar\nu}^A}{d \Omega' d E'} 
= \frac{{G_F}^2}{(2\pi)^2} \; \frac{|{\bf k}^\prime|}{|{\bf k}|} \;
\left(\frac{m_W^2}{q^2-m_W^2}\right)^2
L^{\alpha \beta}_{\nu, \bar\nu}
\; W_{\alpha \beta}^{A}\,,
\end{equation}
where $L^{\alpha \beta}$ is the leptonic tensor and is given by
\begin{equation} 	\label{lepton}
L^{\alpha \beta}=k^{\alpha}k'^{\beta}+k^{\beta}k'^{\alpha}
-k.k^\prime g^{\alpha \beta} \pm i \epsilon^{\alpha \beta \rho \sigma} k_{\rho} 
k'_{\sigma}\,,
\end{equation}
where  plus sign is for antineutrino and minus sign is for neutrino.

$W_{\alpha \beta}^A$ is the nuclear hadronic tensor defined in terms of nuclear 
hadronic structure functions $W^A_{i}(x,Q^2)$: 

\begin{eqnarray}\label{had_ten}
W^A_{\alpha \beta}&=&
\left( \frac{q_{\alpha} q_{\beta}}{q^2} - g_{\alpha \beta} \right) \;
W_1^A
+ \frac{1}{M_A^2}\left( P_{A\alpha} - \frac{P_A . q}{q^2} \; q_{\alpha} \right)
\left( P_{A\beta} - \frac{P_A . q}{q^2} \; q_{\beta} \right)
W_2^A \nonumber \\
&& -\frac{i}{2 M_A^2} \epsilon_{\alpha \beta \rho \sigma} P_{A}^{\rho} q^{\sigma}
W_3^A
\end{eqnarray}
$M_A$ is the mass of the nucleus and $P_A$ is the momentum of the nucleus.  

In the local density approximation the nuclear hadronic tensor $W^A_{\alpha \beta}$ is written as \cite{prc84}:
 \begin{equation}	\label{conv_WA}
W^A_{\alpha \beta} = 4 \int \, d^3 r \, \int \frac{d^3 p}{(2 \pi)^3} \, 
 \int^{\mu}_{- \infty} d p^0 \frac{M}{E ({\bf p})} S_h (p^0, {\bf p}, \rho(r))
W^N_{\alpha \beta} (p, q), 
\end{equation}
$S_h$ is the spectral hole function and has been taken from Ref.~\cite{oset1992}. $\rho(r)$ is the baryon density for the nucleus.
 M and E  is the mass and energy of the nucleon.
 
In the above expression  $W^N_{\alpha \beta} (p, q)$ is the nucleon hadronic tensor and, in the limit of lepton mass $m_l \rightarrow 0$, 
only three structure functions contribute to the cross section. Therefore, in this limit, we may write:
\begin{eqnarray}\label{hadten_nucleon}
W^N_{\alpha \beta} &=&
\left( \frac{q_{\alpha} q_{\beta}}{q^2} - g_{\alpha \beta} \right) \;
W_1^{\nu(\bar \nu)}
+ \frac{1}{M^2}\left( p_{\alpha} - \frac{p . q}{q^2} \; q_{\alpha} \right)
\left( p_{\beta} - \frac{p . q}{q^2} \; q_{\beta} \right)
W_2^{\nu(\bar \nu)} \nonumber \\
&& -\frac{i}{2 M^2} \epsilon_{\alpha \beta \rho \sigma} p^{\rho} q^{\sigma}
W_3^{\nu(\bar \nu)}
\end{eqnarray}
where $W_i^{N}$ are the structure functions, which depend
on the scalars $q^2$ and $p.q$. 

$W^A_i(x,Q^2)$ (i=1-3) given in Eq.(\ref{had_ten}) are redefined in terms of dimensionless structure functions $F^A_{i}(x,Q^2)$ through

\begin{eqnarray}\label{relation}
M_A W_1^A(\nu, Q^2)=F_1^A(x, Q^2),~~~~ \nu W_2^A(\nu, Q^2)=F_2^A(x, Q^2),~~~~\nu W_3^A(\nu, Q^2)=F_3^A(x, Q^2).
\end{eqnarray}

The expressions for $F^A_2$(x,$Q^2$) and $F_3^A$(x,$Q^2$) in the nuclear medium are obtained as \cite{prc84}:

\begin{eqnarray} \label{f2a}
F^A_2(x_A,Q^2)&=&4\int d^3r\int \frac{d^3p}{(2\pi)^3}\frac{M}{E({\bf p})}\int_{-\infty}^\mu dp^0\;
S_h(p^0,\mathbf{p},\rho(\mathbf{r})) \frac{x}{x_N}
\left( 1+\frac{2x_N p_x^2}{M\nu_N} \right)  F_2^N(x_N,Q^2)
\end{eqnarray}
\begin{eqnarray}\label{f3a}
F_3^A(x_A,Q^2)&=&4\int d^3r \int \frac{d^3p}{(2\pi)^3} \frac{M}{E({\bf p})}\int_{-\infty}^{\mu} dp^0
S_h(p^0,\mathbf{p},\rho(\mathbf{r})) \frac{p^0\gamma-p_z}{(p^0-p_z\gamma)\gamma} F_3^N(x_N,Q^2)
\end{eqnarray}

where $F_2^N$ and $F_3^N$ are the nucleon structure functions written in terms of parton distribution functions,
q is the four momentum transfer, $q^2 = -Q^2$, x is the Bjorken variable, $\nu$ is the energy transfer,

\begin{eqnarray*}	
\gamma=\frac{q_z}{q^0}=\left(1+\frac{4M^2x^2}{Q^2}\right)^{1/2}\,~\rm{and} ~~x_N=\frac{Q^2}{2(p^0q^0-p_zq_z)}.
\end{eqnarray*}
\begin{figure}
  \includegraphics[height=.3\textheight,width=14cm]{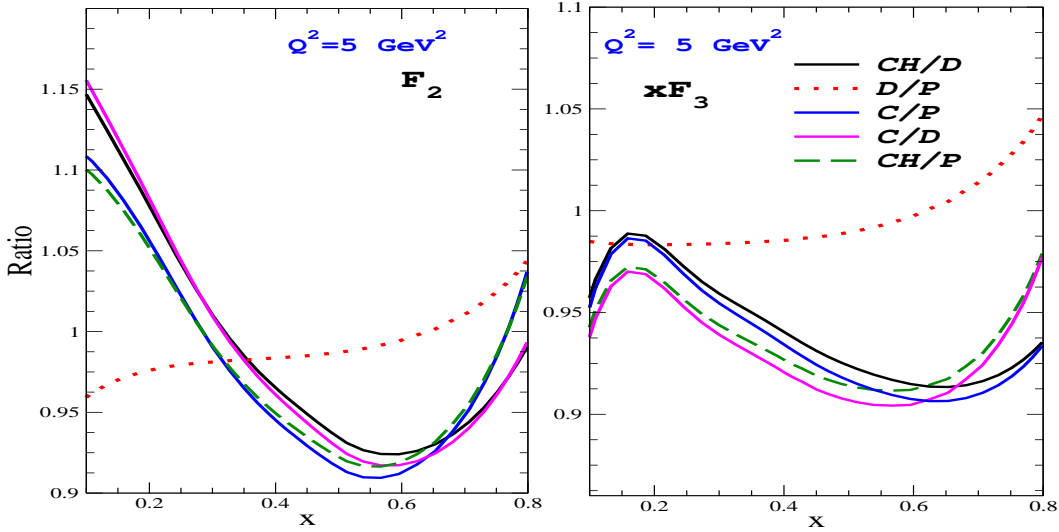}
  \caption{Ratio $R(x,Q^2)=F_i^{CH}/F_i^P, F_i^{CH}/F_i^D, F_i^D/F_i^P, F_i^C/F_i^P, F_i^C/F_i^D$(i=2(left) and 3(right))  using full model at NLO.}
\label{fig1}
\end{figure} 
  All the nuclear information like Pauli blocking, Fermi motion, nucleon correlation are contained in the spectral function. The results obtained by using Eqs.(\ref{f2a}) and (\ref{f3a}) with
 target  mass  correction is our base result. When shadowing and anti-shadowing effects for $F_2^A$ and
 $F_3^A$, as well as pion and rho cloud contributions for $F_2^A$, are taken into account, results obtained by including all these effects, we call this as the results with full calculations. 
Using the results of weak structure functions $F_2^A$ and  $F_3^A$, we have obtained the ratio of structure
 functions $F_i^A/F_i^{proton}$ and $F_i^A/F_i^{deuteron}$. For the numerical calculations, parton distribution functions
for the nucleons have been taken from the parametrization
of the Coordinated Theoretical-Experimental Project on QCD
(CTEQ) Collaboration (CTEQ6.6)~\cite{cteq}. The Next-to-Leading-
Order (NLO) evolution of the deep inelastic structure functions
has been taken from the works of Vermaseren et al.~\cite{moch}.

\section{Results and Discussion}
\begin{figure}
\includegraphics[height=.3\textheight,,width=14cm]{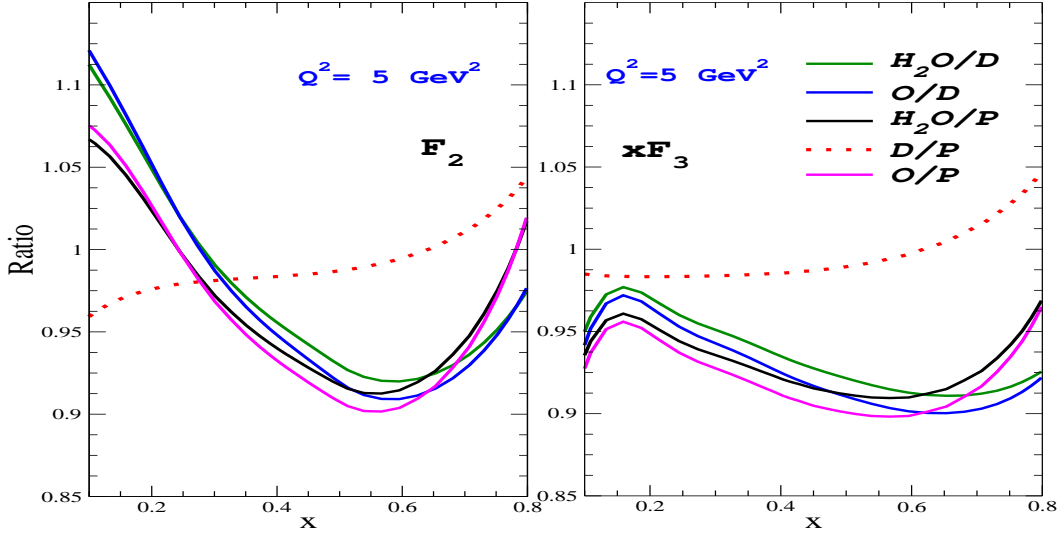}
\caption{Ratio $R(x,Q^2)=F_i^{H_{2}O}/F_i^P, F_i^{H_{2}O}/F_i^D, F_i^D/F_i^P, F_i^O/F_i^P, F_i^O/F_i^D$(i=2(left) and 3(right))using full model at NLO.}
\label{fig2}
\end{figure}

We have studied the medium effects on the weak structure functions $F_2^A$ and $F_3^A$ in the charged current anti(neutrino) induced deep inelastic reactions
using carbon and oxygen targets for different values of x at the next-to-leading order(NLO). We find that the difference 
between the base results and the full calculations is about 10-12$\%$ and 4-6$\%$ at low values of x in the case of $F_2^A$ and $F_3^A$ respectively and this difference
reduces to 1$\%$ at higher x for both the cases. We have presented the results for the ratio $F_i^A/F_i^{proton}$ and $F_i^A/F_i^{deuteron}$(i=2,3) in
Figs.~\ref{fig1}-\ref{fig2}.
and observe that the nuclear medium effects are not the same in different nuclear targets, as well as
the nature of medium effects in $F_2^A$ and  $F_3^A$ are different.
Furthermore, the ratio of $F_i^{deuteron}/F_i^{proton}$ (i=2,3) has also been shown in Figs.~\ref{fig1}-\ref{fig2} and we find that the medium effects in deuteron are 
quite different from the medium effects in heavy nuclei. This study may be useful in the analysis  of MINER$\nu$A and other proposed anti(neutrino) experiments 
in the energy region of few GeV using nuclear targets.

%%%%%%%%%%%%%%%%%%%%%%%%%%%%%%%%%%%%%%%%%%%%
%% Sample figure:
%%
%% The option [height=...] scales the picture to the given height,
%% without it it would be printed at its nominal size
%%%%%%%%%%%%%%%%%%%%%%%%%%%%%%%%%%%%%%%%%%%%

%%%%%%%%%%%%%%%%%%%%%%%%%%%%%%%%%%%%%%%%%%%%%%%%
%% BACKMATTER
%%%%%%%%%%%%%%%%%%%%%%%%%%%%%%%%%%%%%%%%%%%%%%%%

\section*{ACKNOWLEDGMENTS}
  The authors thank M. J. Vicente Vacas, University of Valencia, Spain for many useful discussions and encouragement throughout this work.

%%%%%%%%%%%%%%%%%%%%%%%%%%%%%%%%%%%%%%%%%%%%%%%%
%% The bibliography can be prepared using the BibTeX program or
%% manually.
%%
%% The code below assumes that BibTeX is used.  If the bibliography is
%% produced without BibTeX comment out the following lines and see the
%% aipguide.pdf for further information.
%%
%% For your convenience a manually coded example is appended
%% after the \end{document}
%%%%%%%%%%%%%%%%%%%%%%%%%%%%%%%%%%%%%%%%%%%%%%%%

%%%%%%%%%%%%%%%%%%%%%%%%%%%%%%%%%%%%%%%%%%%%%%%%
%% You may have to change the BibTeX style below, depending on your
%% setup or preferences.
%%
%%
%% For The AIP proceedings layouts use either
%%%%%%%%%%%%%%%%%%%%%%%%%%%%%%%%%%%%%%%%%%%%

\end{document}